\documentclass[sn-aps]{sn-jnl}
\usepackage{braket}
\usepackage{array}
\usepackage{dsfont}
\usepackage{amsmath}
\usepackage{pifont}
\usepackage[normalem]{ulem}
\usepackage{amssymb}
\usepackage{xcolor}
\usepackage{amsfonts}
\usepackage{mathtools}
\usepackage{graphicx}
\usepackage{dcolumn}
\usepackage{bm}
\usepackage{multirow}
\usepackage{hhline}

\usepackage{leftidx}
\usepackage{color}
\usepackage{mathtools}
\usepackage{MnSymbol}
\usepackage[mathscr]{eucal}
\usepackage[capitalize]{cleveref}

\usepackage[utf8]{inputenc}
\usepackage{natbib}
\usepackage{comment}
\usepackage{xcolor}
\newcommand{\abs}[1]{\left| #1 \right|}

\usepackage[normalem]{ulem}

\begin{document}

\title[Article Title]{Density effects on the interferometry of Efimov states by modulating magnetic fields}

	\author*[1]{\fnm{G.} \sur{Bougas}}\email{gbougas@mst.edu}

\author[1]{\fnm{S. I.} \sur{Mistakidis}}\email{smystakidis@mst.edu}

\author[2]{\fnm{P.} \sur{Giannakeas}}\email{pgiannak@pks.mpg.de}

\affil*[1]{\orgdiv{Physics Department}, \orgname{Missouri University of Science and Technology}, \orgaddress{ \city{Rolla, MO}, \postcode{65409}, \country{USA}}}

\affil[2]{\orgname{Max-Planck-Institut f\"ur Physik komplexer Systeme}, \orgaddress{\street{Nöthnitzer Str. 38}, \postcode{ D-01187}, \city{Dresden}, \country{Germany}}}

\abstract{
Dynamical association of Efimov trimers in thermal gases by means of modulated magnetic fields has proven very fruitful in determining the binding energy of trimers. 
The latter was extracted from the number of remaining atoms, which featured oscillatory fringes stemming from the superposition of trimers with atom-dimers. 
Subsequent theoretical investigations utilizing the time-dependent three-body problem revealed additional association mechanisms, manifested as superpositions of the Efimov state with the trap states and the latter with atom-dimers. 
The three atoms were initialized in a way to emulate a thermal gas with uniform density. 
Here, this analysis is extended by taking into account the effects of the density profile of a semi-classical thermal gas. 
The supersposition of the Efimov trimer with the first atom-dimer remains the same, while the frequencies of highly oscillatory fringes shift to lower values. 
The latter refer to the frequencies of trimers and atom-dimers in free space since the density profile smears out the contribution of trap states.}

\maketitle
	
\section{Introduction}\label{Sec:Intro}
 
Dilute ultracold gaseous matter constitutes the main platform in order to explore unique and exotic attributes of few-body systems in a controllable manner.
For example, ultracold experiments achieved to shed light on the existence of Efimovian triatomic molecules \cite{kraemer_evidence_2006}; an exotic three-body system which was originally perceived 
by V. Efimov in the context of nuclear physics \cite{Efimov_energy_1970,Efimov_energy_1973,Efimov_weakly_1971}.
Specifically, it was theoretically predicted that an infinite geometric progression of three-body bound states emerges from unbound pairwise two-body subsystems \cite{greene_universal_2017,nielsen_three-body_2001,naidon_efimov_2017,dincao_few-body_2018}.

Recent theoretical \cite{Kievsky_efimov_2021,riisager_nuclear_1994,petrov_three-body_2015,Sun_Hermitian_2023,giannakeasprl2018,braaten_universality_2006,blume_few-body_2012,wang_chapter_2013,naidon_microscopic_2014,naidon_physical_2014,giannakeasfewbody2016,mestromEfimovVanWaals2017,mestromVanWaalsUniversality2020}  and experimental \cite{Kunitski_observation_2015,pires_observation_2014prl, ulmanis_universality_2015,hafner2017role,tung_geometric_2014-1,ferlaino_forty_2010,gross_observation_2009,roy_prl_2013,johansen_testing_2017} advances provide an in depth exploration of the properties and the universal idiosyncrasies of the Efimovian physics in the realm of nuclear and atomic physics.
For instance, ultracold experiments that use radio-frequency spectroscopy or magnetoassociation techniques have accurately measured the binding energy of Efimov states \cite{Lompe_radio_2010,Nakajima_measurement_2011,Machtey_association_2012} or interaction quench/sweeps in order to probe their lifetime \cite{klauss_observation_2017}.
Yudkin {\it et al.} introduced a Ramsey-type protocol where a thermal gas is exposed to a double sequence of modulated magnetic fields, i.e. magnetic field pulses, for variable in between times  \cite{yudkin_coherent_2019,yudkin_reshaped_2024}.
This Ramsey-type interferometer allowed to probe the challenging regime below the atom-dimer threshold without using a unitary Bose gas, accurately measuring the trimer binding energies.
However, an overall damping in the oscillations of the number of remaining atoms was observed, where the corresponding decay time scale exceeds the typical lifetime of Efimov trimers even for $^{85}$Rb$_3$ \cite{klauss_observation_2017}.
Theoretical investigations demonstrated that this Ramsey-type protocol possesses a general scope and can be used to probe Efimov states not only near the atom-dimer threshold, but also below the break-up threshold \cite{Bougas_interferometry_2023}.
Furthermore, Ref.~\cite{Bougas_interferometry_2023} explicated in the case of thermal gases that the decay time of interferometric fringes is twice as large as the lifetime of the Efimov states. 
This means that exposing gases in magnetic field pulses permits us to simultaneously assess the binding energies and lifetimes of Efimov states.

In this work, we further investigate theoretically the coherent association of Efimov trimers in a spherical trap, which is induced by a sequence of double magnetic field pulses emphasizing the role of a thermal gas density component.
More specifically, in Ref. \cite{Bougas_interferometry_2023} in the corresponding three-body calculations the trapping frequency was rescaled by the peak density of a thermal gas of $^{85}$Rb atoms. 
This means the density dependent trapping frequency ensures that the few-body system exhibits the same diluteness as in a standard experimental scenario.
Here, we consider the same system as in Ref. \cite{Bougas_interferometry_2023} taking into account a more realistic density profile, therefore going beyond previous studies.
In particular, the physically relevant quantity is the thermally averaged probability to occupy an Efimov state which is integrated over a profile of density dependent trapping frequencies.

Our analysis shows that the averaged probabilities still exhibit signatures of coherence by varying the time between the two magnetic field pulses, i.e. the dark time.
Indeed, the corresponding Fourier spectrum of the thermally averaged probability to occupy trimers possesses predominantly three frequencies: (i) two high ones, corresponding to the energy difference of the Efimov trimer with the break-up threshold, and the latter with the atom-dimer states, and (ii) a low frequency which originates from the energy difference between the Efimov and atom-dimer states.
Comparing these results with the predictions of  Ref.~\cite{Bougas_interferometry_2023} we conclude that the density profile of a thermal gas only affects the high frequencies which approach those of a three-body system without the spherical trap.
Furthermore, the low frequency is independent of the density implying that the Ramsey-type dynamical protocol can be utilized to probe few-body state-to-state coherences without being smeared out by density and thermal effects.

This work is structured as follows. In Sec. \ref{Sec:Framework} we introduce the characteristics of the time-dependent three-body systems with trapping frequencies associated to the density profile of a thermal gas. In Sec. \ref{Sec:Density_effects} we discuss the impact of the density profile of a thermal gas on the thermally averaged probability to occupy trimers. Finally in Sec. \ref{Sec:Conclusions} we summarize our results and discuss further perspectives.

\section{Time-Dependent three-body system and correspondence to a thermal gas}  \label{Sec:Framework}
Owing to the dilutness of thermal gases, one can study density effects on three-body harmonically trapped systems by utilizing density dependent confinement frequencies~\cite{borca_two-atom_2003,dincao_efimov_2018}. 
As such, the density of the few-body system matches the one of a condensate or a thermal gas.
Therefore, as a paradigm in the following we focus on the dynamics of three $^{85}$Rb atoms of mass $m$.
Importantly, we consider an ensemble of three-body systems, each trapped in a radially symmetric trap with frequency $\omega^{(\mathbf{j})}_r$ where the index $(\mathbf{j})$ labels the setup. 
The frequencies are chosen such that the harmonic oscillator length $a^{(\mathbf{j})}_r = \sqrt{ \hbar / (m \omega^{(\mathbf{j})}_r)}$ corresponds to a mean interparticle distance, $a^{(\mathbf{j})}_r \simeq [n^{(\mathbf{j})}]^{-1/3} $~\cite{sykes_quenching_2014,borca_two-atom_2003,dincao_efimov_2018} between atoms in the thermal gas. The $n^{(\mathbf{j})}$ 
span the one-body density profile of a non-interacting semi-classical gas trapped in a harmonic trap with radial frequency $\Omega_r = 2\pi \times 10 \, \rm{Hz}$ and peak density at the trap center $n_0=5 \times 10^{12} \, \rm{cm}^3$, following the experiment of Ref.~\cite{klauss_observation_2017}.
In particular, the assumed densities lie within the interval $[0.6,1]n_0$, and the trap frequency associated to the largest (smallest) density is denoted by $\omega^{(\mathbf{1})}_r$ ($\omega^{(\mathbf{c})}_r$). 
The one-body density profile reads explicitly~\cite{pethick_bose_2008},
\begin{equation}
n(\boldsymbol{x}) = n_0 \exp  \{ -\frac{1}{2} m \Omega_r^2 \boldsymbol{x}^2 / (k_B \mathcal{T})  \},
\label{Eq:Semiclassical}
\end{equation}
where $\mathcal{T}$ is the temperature, $k_B$ is the Boltzman constant, and $\boldsymbol{x}$ are the three-dimensional coordinates of the thermal cloud. 
The above expression relates the density $n^{(\mathbf{j})}$ and the spatial coordinates of the thermal cloud, i.e. $n^{(\mathbf{j})}=n(\boldsymbol{x}^{(\mathbf{j})})$. 
	
More concretely, for every trapping frequency $\omega^{(\mathbf{j})}_r \simeq \hbar [n^{(\mathbf{j})}]^{2/3}/m$, we solve the time-dependent three-body Schr\"odinger equation pertaining to the below given Hamiltonian featuring time-dependent $s$-wave scattering length $a(t)$~\cite{Bougas_interferometry_2023}. Namely,  
\begin{gather}
\mathcal{H}^{(\mathbf{j})}(t)=\sum_{i=1}^{3} \left( \frac{-\hbar^2  \nabla_i^2}{2m}  +\frac{m [\omega_r^{(\mathbf{j})} ]^2}{2} \boldsymbol{r}_i^2 \right)  +\sum_{i<j} \frac{4\pi \hbar^2 a(t)}{m} \delta^{(3)}(\boldsymbol{r}_{ij})  \left( \partial_{r_{ij}}  r_{ij} \cdot \right),
\label{Eq:hamilt_lab}
\end{gather}  
where $r_{ij} = \abs{\boldsymbol{r}_i - \boldsymbol{r}_j}$, and $\boldsymbol{r}_i$ being the position of the $i$-th indistinguishable particle. The $s$-wave scattering length,  $a(t)$, is modulated according to the following experimentally realizable~\cite{yudkin_coherent_2019,yudkin_reshaped_2024,Bougas_interferometry_2023} double pulse Ramsey-type sequence 
\begin{gather}
a(t)=  a_{bg}+a_m \cos{(\Omega t)} \Big[   \chi(t)+\chi(t-t_d-2t_0-\tau)  \Big], \label{Eq:Pulse} \\
\rm{with~envelope}~~\chi(t)= \begin{cases}   \sin^2\left( \frac{\pi t}{2t_0} \right), & 0\leq t < t_0\\
1, & t_0 \leq t < t_0+\tau \\
\sin^2\left( \frac{\pi (t-\tau)}{2t_0} \right), & t_0+\tau \leq t \leq 2t_0+\tau \\
0, & \rm{otherwise}
\end{cases}. \label{Eq:Pulse_envelope}
\end{gather}
It becomes evident that the instantaneous scattering length $a(t)$ oscillates around the background scattering length $a_{bg}$ with driving amplitude $a_m \sim 0.2 \, a_{bg}$, and 
frequency $\Omega$.
The two pulses are characterized by a time-dependent envelope $\chi(t)$, where $t_0$ and $\tau$ signify the ramp on/off times and length of the pulse envelope, respectively. 
The pulse characteristics, which will be explicated in what follows, are designed such that Efimov trimers and energetically low atom dimers are coherently populated~\cite{Bougas_interferometry_2023,yudkin_coherent_2019,yudkin_reshaped_2024}. 
During the {\it dark time} $t_d$, referring to the time-interval in  between the two pulses, the three-body system evolves freely.

\begin{figure}[t!]
\centering
\includegraphics[width=0.6 \textwidth]{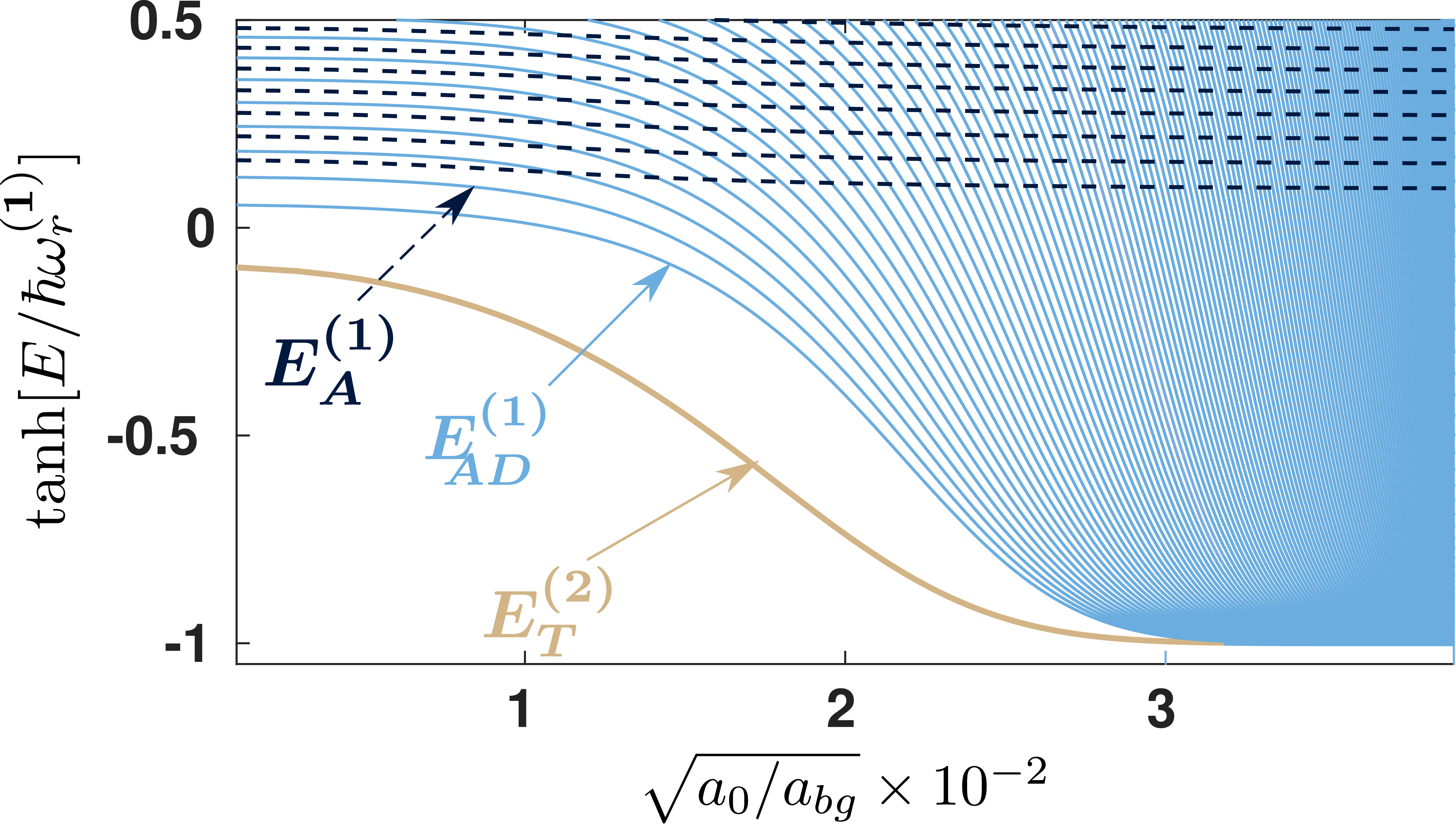}
\caption{Rescaled energy spectrum of three atoms in a trap with radial frequency $\omega^{(\mathbf{1})}_r$, corresponding to the peak density $n_0$. The eigenstates fall into three categories, trimers (T), atom-dimers (AD) and three atom states (A).}
\label{Fig:Spectrum}
\end{figure}

To solve Eq.~\eqref{Eq:hamilt_lab}, we transform the above-described three-body Hamiltonian from the lab to the center-of-mass frame, $\mathcal{H}^{(\mathbf{j})}(t) = \mathcal{H}^{(\mathbf{j})}_{\rm{cm}} + \mathcal{H}^{(\mathbf{j})}_{\rm{rel}}(t)$. Here,  only the relative Hamiltonian $\mathcal{H}^{(\mathbf{j})}_{\rm{rel}}(t)$ depends on the time-dependent scattering length. Subsequently, a change from cartesian to hyperspherical coordinates, see also  Refs.~\cite{greene_universal_2017,nielsen_three-body_2001,naidon_efimov_2017} for technical details, is performed. In this context, the relative degrees of freedom of the system consist of the hyperradius $R$ that controls the overall system size, and the five hyperangles $\boldsymbol{\varpi}$, which determine the relative distance between the three atoms. When cast in this coordinate frame, the relative Hamiltonian reads
\begin{gather}
\mathcal{H}^{(\mathbf{j})}_{\rm{rel}}(t) = -\frac{\hbar^2}{2\mu} \frac{1}{R^{5/2}}  
\frac{\partial^2}{\partial R^2} \left( R^{5/2} \cdot    \right) +\frac{15 \hbar^2}{8 \mu R^2} +\frac{\hbar^2 \boldsymbol{\Lambda}^2}{2\mu R^2} \nonumber \\
+\frac{1}{2} \mu [\omega^{(\mathbf{j})}_r]^2 R^2 
+a_{bg} V(R;\boldsymbol{\varpi}) +V(R;\boldsymbol{\varpi}) [a(t)-a_{bg}] \nonumber \\
= \mathcal{H}^{(\mathbf{j})}_{bg}(R;\boldsymbol{\varpi}) + a_m V(R;\boldsymbol{\varpi}) \cos{(\Omega t)} \Big[   \chi(t)+\chi(t-t_d-2t_0-\tau)  \Big],
\label{Eq:Rel_Hamilt_Hyper}
\end{gather}
where $\mu=m/\sqrt{3}$ is the three-body reduced mass and $\boldsymbol{\Lambda}^2$ is the grand angular momentum operator describing the total angular momentum of three atoms~\cite{avery_hyperspherical_1989}. Also, the $V(R;\boldsymbol{\varpi})$ term is the regularized contact interaction potential expressed in hyperspherical coordinates. Note that we have set apart the time-dependence in the last term of Eq. \eqref{Eq:Rel_Hamilt_Hyper}. 
 
To subsequently tackle the time-dependent relative Hamiltonian, we employ the basis of the field-free Hamiltonian $\mathcal{H}^{(\mathbf{j})}_{bg}$. The latter is diagonalized within the adiabatic hyperspherical approach~\cite{rittenhouse_greens_2010,naidon_efimov_2017,greene_universal_2017,nielsen_three-body_2001,dincao_few-body_2018}, yielding the field-free eigenstates, $\{ \ket{n^{(\mathbf{j})}} \}$ with corresponding energy $E^{(\mathbf{j})}_n$. The main idea behind this approach is to write the eigenstates in a basis of hyperangular functions $\Phi^{(\mathbf{j})}_{\nu}(R;\boldsymbol{\varpi})$, treating $R$ as an adiabatic parameter, 
\begin{equation}
\braket{R,\boldsymbol{\varpi} | n^{(\mathbf{j})}} = R^{-5/2} \sum_{\nu} F^{(n,\mathbf{j})}_{\nu}(R) \Phi^{(\mathbf{j})}_{\nu}(R;\boldsymbol{\varpi}),
\label{Eq:Adiabatic_expansion}
\end{equation} 
where $F^{(n,\mathbf{j})}_{\nu}(R)$ are the so-called hyperradial channels. 
The $\Phi^{(\mathbf{j})}_{\nu}(R;\boldsymbol{\varpi})$ basis functions are obtained by solving the field-free relative Hamiltonian at fixed hyperradius $R$, i.e. omitting the derivatives, 
\begin{equation}
\mathcal{H}^{(\mathbf{j})}_{bg}(R=const;\boldsymbol{\varpi}) \Phi^{(\mathbf{j})}_{\nu}(R=const;\boldsymbol{\varpi}) = U^{(\mathbf{j})}_{\nu}(R=const) \Phi^{(\mathbf{j})}_{\nu}(R=const;\boldsymbol{\varpi}),
\label{Eq:Potential_curves}
\end{equation}
resulting in a set of potential curves, $U^{(\mathbf{j})}_{\nu}(R)$. Subsequently, plugging in the expansion \eqref{Eq:Adiabatic_expansion} into the field-free Hamiltonian and integrating over the hyperangles $\boldsymbol{\varpi}$, a set of coupled differential equations~\cite{greene_universal_2017} is obtained for the hyperradial channels $F^{(n,\mathbf{j})}_{\nu}(R)$,
\begin{gather}
-\frac{\hbar^2}{2\mu} \frac{d^2}{dR^2} F^{(n,\mathbf{j})}_{\nu}(R) + U^{(\mathbf{j})}_{\nu}(R)  F^{(n,\mathbf{j})}_{\nu}(R) \nonumber \\
-\frac{\hbar^2}{2\mu} \sum_{\nu'} \Bigg[ \braket{\Phi^{(\mathbf{j})}_{\nu}(R;\boldsymbol{\varpi}) | \frac{\partial^2 }{\partial R^2} \Phi^{(\mathbf{j})}_{\nu'}(R;\boldsymbol{\varpi}) }_{\boldsymbol{\varpi}}  F^{(n,\mathbf{j})}_{\nu'}(R) \nonumber \\
+ 2\braket{\Phi^{(\mathbf{j})}_{\nu}(R;\boldsymbol{\varpi}) | \frac{\partial }{\partial R} \Phi^{(\mathbf{j})}_{\nu'}(R;\boldsymbol{\varpi}) }_{\boldsymbol{\varpi}}  \frac{d}{d R} F^{(n,\mathbf{j})}_{\nu'}(R)   \Bigg] = E^{(\mathbf{j})}_n F^{(n,\mathbf{j})}_{\nu}(R),
\label{Eq:Hyperradial_equations}
\end{gather}
where the braket notation, $\braket{}_{\boldsymbol{\varpi}}$, refers to the integration only over the hyperangles.

The resulting field-free eigenstates are categorized into three classes as can be readily seen in the three-body spectrum depicted in Fig.~\ref{Fig:Spectrum} as a function of the $s$-wave scattering length~\cite{bougas2021few}. Namely, Efimov trimers (T) (orange solid line in Fig. \ref{Fig:Spectrum}), atom-dimers (AD) (blue solid lines) and three-atom states (A) (black dashed lines). 
For positive $a_{bg}$, the excited Efimov trimer approaches the atom-dimer states and eventually dissociates~\cite{giannakeasfewbody2016,braaten_universality_2006}. 
This interaction regime was employed in~\cite{yudkin_coherent_2019,yudkin_reshaped_2024} in order to create coherent superposition of the first excited Efimov trimer with the first atom-dimer. 
Apart from the driving frequency $\Omega = \abs{E^{(1)}_{AD}-E^{(1)}_A}/h$, the pulse envelope is chosen such that both the first excited Efimov state and the first atom-dimer are not energetically resolved by the pulse~\cite{yudkin_coherent_2019}. 
This condition refers to the full-width-at-half-maximum of the Fourier transform of $\chi(t)$ being comparable to $\abs{E^{(2)}_T-E^{(1)}_{AD}}/h$. 
Note that the driving frequency and the pulse characteristics depend on the considered trapping frequency $\omega_r^{(\mathbf{j})}$. 
However, these parameters do not strongly vary since the Efimov trimer and first atom-dimer states are energetically deep, and hence they are not affected by the trap. 
In contrast, the three-atom and highly-excited atom-dimer states are the only ones that are strongly influenced by the trap frequency.
     
\begin{figure}[t]
\centering
\includegraphics[width=1 \textwidth]{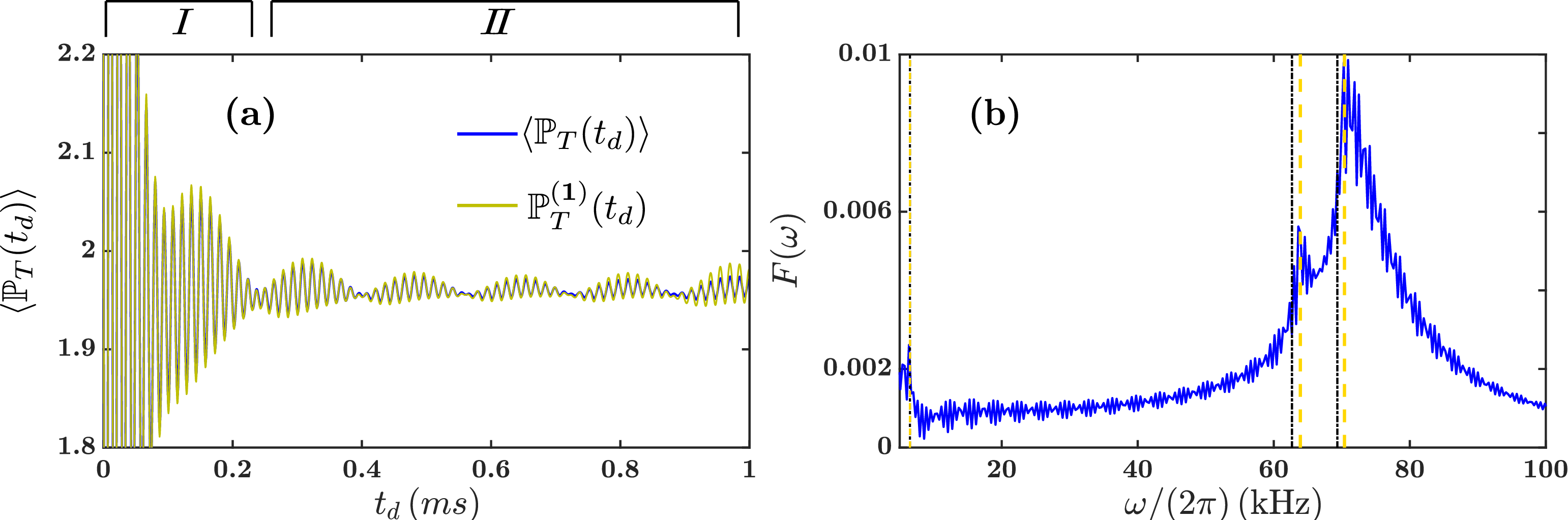}
\caption{(a) Comparison of the ratio of thermally averaged probabilities taking into account a realistic density profile for the thermal gas (blue line), or just the peak density (yellow line) at $\mathcal{T}=270 \, \rm{nK}$. High-amplitude oscillations appear in region I, which are subsequently damped in region II due to thermal effects. (b) Frequency spectrum stemming from $\langle \mathbb{P}_T(t_d) \rangle$. The low-frequency peak (black dashed line) remains unchanged regardless of whether we consider the realistic density profile or solely  the peak density. The other two high frequencies are slightly shifted from the corresponding values with just the peak density (yellow dashed lines), towards frequencies corresponding to the first atom-dimer and first excited Efimov state in the absence of any trap (black dashed-dotted lines).} 
\label{Fig:Density_average}
\end{figure}

For every trap frequency, $\omega_r^{(\mathbf{j})}$, the three-body system is initialized in an ensemble of three-atom states, populated according to the Maxwell-Boltzmann distribution at temperature $\mathcal{T}$. In this way, thermal effects are taken into account~\cite{yan_harmonically_2013,giannakeas_nonadiabatic_2019}.
Subsequently, the time-dependent relative wavefunction assigned to an initial three atom-state $\alpha$  is expanded in the field-free basis, $\ket{\Psi^{(\alpha,\mathbf{j})}_{\rm{rel}}(t)} = \sum_n c^{(\alpha,\mathbf{j})}_n(t) \ket{n^{(\mathbf{j})}}$, where $c^{(\alpha,\mathbf{j})}_n(t)$ is the probability amplitude of the $n$-th stationary state.

Plugging the above expression into the time-dependent Schr\"odinger equation (TDSE) results in a matrix differential equation for the probability amplitudes~\cite{Bougas_interferometry_2023},

\begin{equation}
i \hbar \frac{d \boldsymbol{c}^{(\alpha,\mathbf{j})}(t)}{dt} = \boldsymbol{\mathcal{H}}^{(\mathbf{j})}_{rel}(t) \cdot \boldsymbol{c}^{(\alpha,\mathbf{j})}(t),
\label{Eq:Expansion_coefficients}
\end{equation}
where $\boldsymbol{\mathcal{H}}^{(\mathbf{j})}_{rel}(t)$ is the relative Hamiltonian matrix expressed in the field-free basis. A second-order split-operator method is utilized to solve the above equation~\cite{burstein_third_1970}.

Since the aim is to monitor
the dynamical population of Efimov trimers in a thermal gas, we introduce the ratio of the thermally averaged probability to occupy trimers after the second pulse (nominator in Eq.~(\ref{Eq:Efimov_fraction})) with respect to the end of the first pulse (denominator in Eq.~(\ref{Eq:Efimov_fraction})). Specifically, this observable takes the form
\begin{subequations}
\begin{gather}
\mathbb{P}^{(\mathbf{j})}_{T}(t_d)=\frac{\sum_{\alpha \in A} \sum_{k \in \, T} e^{-\frac{E^{(\mathbf{j})}_{\alpha}}{k_B \mathcal{T}}} \abs{c^{(\alpha,\mathbf{j})}_k(2\tilde{\tau}+t_d)}^2}{\sum_{\alpha \in A} \sum_{k \in \, T}  e^{-\frac{E_{\alpha}^{(\mathbf{j})}}{k_B \mathcal{T}}} \abs{c^{(\alpha,\mathbf{j})}_k(\tilde{\tau})}^2 }, \label{Eq:Efimov_fraction} \\
c^{(\alpha,\mathbf{j})}_k(2\tilde{\tau}+t_d)=\sum_{n} U^{(\mathbf{j})}_{kn}(2\tilde{\tau}+t_d,\tilde{\tau}+t_d) e^{-i E^{(\mathbf{j})}_n t_d/\hbar} U^{(\mathbf{j})}_{n \alpha}(\tilde{\tau},0).
\label{Eq:Evolution_op}
\end{gather}
\end{subequations}
In the above expressions, $\tilde{\tau}= 2t_0 + \tau$ is the pulse duration and $U^{(\mathbf{j})}_{i j}(\cdot, \cdot)$ represents the three-body evolution operator (in terms of the field-free basis) during a single pulse.

Next, the thermally averaged quantity in \cref{Eq:Efimov_fraction} is averaged also over the one-body density profile of the thermal gas~\cite{chen_suppression_2022} according to
\begin{equation}
\braket{ \mathbb{P}_T(t_d)  } = \dfrac{  \sum_{\mathbf{j}} \Delta \boldsymbol{x}^{(\mathbf{j})} n(\boldsymbol{x}^{(\mathbf{j})})  \mathbb{P}^{(\mathbf{j})}_T(t_d)   }{    \sum_{\mathbf{j}}   \Delta \boldsymbol{x}^{(\mathbf{j})} n(\boldsymbol{x}^{(\mathbf{j})})    }.
\label{Eq:Density_average}
\end{equation}
It is convenient to perform two variable substitutions $ \abs{ \boldsymbol{x}^{(\mathbf{j})} } \to n^{(\mathbf{j})} \to \omega_r^{(\mathbf{j})}$ 
and express \cref{Eq:Density_average} with respect to the trap frequency yielding the following relation 
\begin{equation}
\braket{  \mathbb{P}_T(t_d) } = \dfrac{  \sum_{\mathbf{j}} \Delta \omega_r^{(\mathbf{j})}  \sqrt{\omega_r^{(\mathbf{j})}} \sqrt{\ln (\omega_r^{(\mathbf{1})} / \omega_r^{(\mathbf{j})}  )  }   \mathbb{P}^{(\mathbf{j})}_T(t_d)   }{  \sum_{\mathbf{j}} \Delta \omega_r^{(\mathbf{j})}  \sqrt{\omega_r^{(\mathbf{j})}} \sqrt{\ln (\omega_r^{(\mathbf{1})} / \omega_r^{(\mathbf{j})}  )  }    }.
\label{Eq:Density_average_final}
\end{equation}
For the above formula the explicit one-body density profile [Eq. \eqref{Eq:Semiclassical}] has been employed, as well as the connection between the trap frequency and the density introduced before Eq. \eqref{Eq:hamilt_lab}.
Importantly,
\cref{Eq:Density_average_final} incorporates the thermal and density effects on the population of Efimovian trimer states induced by the Ramsey-type dynamics through  Eq. \eqref{Eq:hamilt_lab}.

\begin{figure}[t!]
\centering
\includegraphics[width=0.7 \textwidth]{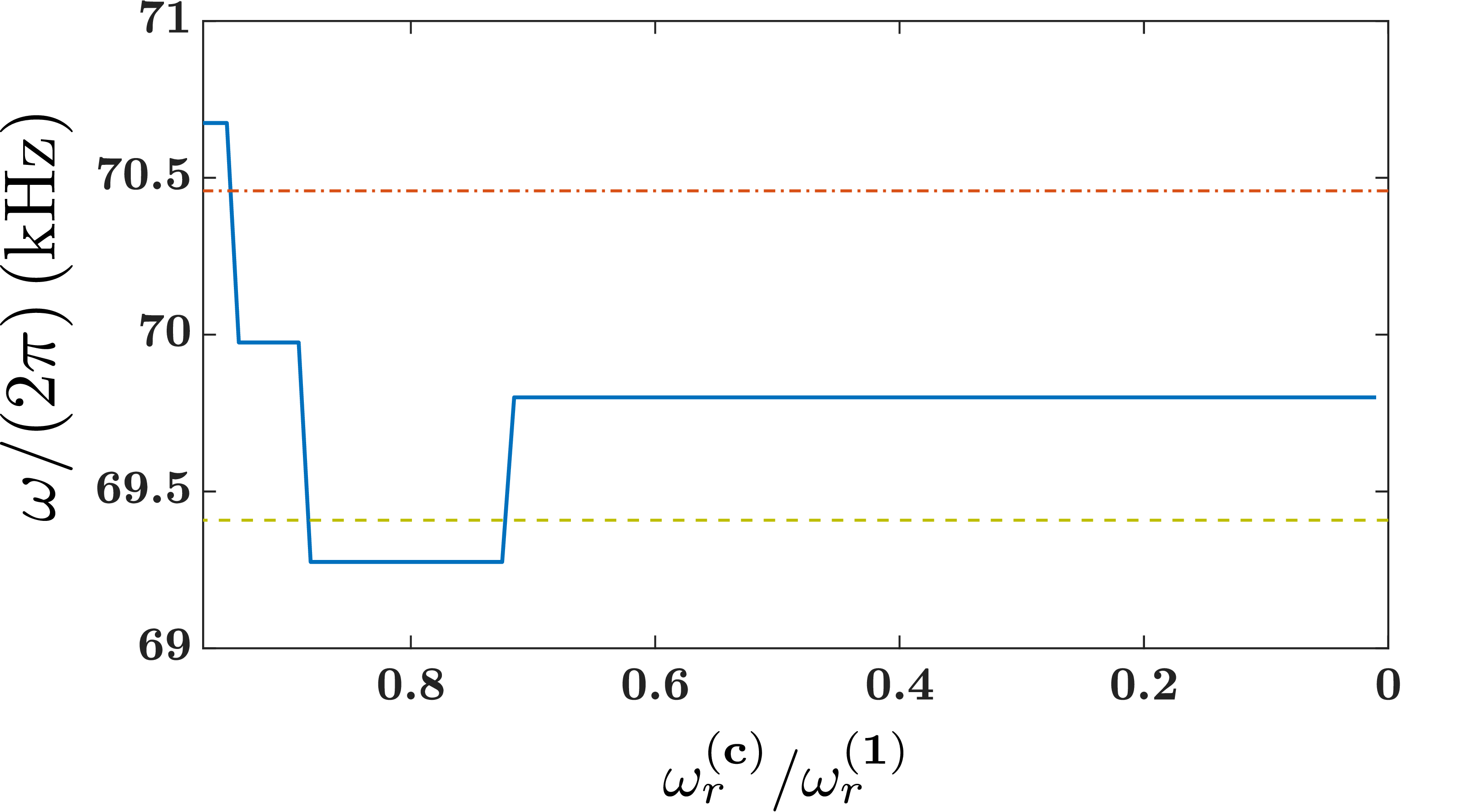}
\caption{Position of the highest frequency peak originating from the thermally averaged probability $\braket{ \mathbb{P}_T(t_d) }$ within the three-level model with respect to the frequency related to the smallest density, $\omega^{(\mathbf{c})}_r$. Note that $\omega^{(\mathbf{1})}_r$ is associated to the peak density. The dash-dotted (dashed) horizontal line corresponds to the energy difference between $E^{(2)}_T$ and $E^{(1)}_A$ ($E^{(2)}_T$ and the energy of the three-atom dissociation threshold in free space).}
    \label{Fig:Cutoff_dependence}
\end{figure}

\section{Effect of the thermal gas density distribution} \label{Sec:Density_effects}

To highlight the impact of the density profile of a thermal gas, the thermally averaged probability of trimers $\braket{ \mathbb{P}_T(t_d) }$ (see \cref{Eq:Density_average_final}) obtained within the full three-body model is compared with the thermally averaged probability ratio pertaining only to $\omega^{(\mathbf{1})}_r$, $\mathbb{P}^{(\mathbf{1})}_T(t_d)$, (see \cref{Eq:Efimov_fraction}) at $\mathcal{T} = 270 \, nK$ [blue versus yellow lines in Fig. \ref{Fig:Density_average}(a)]. 
Clear oscillatory fringes manifest in both probability ratios, and the peak-to-peak amplitude is slightly smaller in the case where the density profile of a thermal gas is taken into account, due to the density average [Eq. \eqref{Eq:Density_average_final}].
In line with the results presented in Ref.~\cite{Bougas_interferometry_2023}, two distinct regimes appear during the free evolution time $t_d$. 
The first one [region $I$ in Fig. \ref{Fig:Density_average}(a)], refers to high-amplitude high-frequency oscillations.
At later dark times the amplitude drops due to thermal effects, revealing low-frequency low-amplitude oscillations 
[region $II$ in Fig. \ref{Fig:Density_average}(a)].

To assess and characterize the participating frequencies, the spectrum of the   $\braket{ \mathbb{P}_T(t_d) }$ signal is presented in Fig. \ref{Fig:Density_average}(b).
To understand the role of the density profile on the dynamical trimer occupations, the three dominant frequencies pertaining to $\mathbb{P}^{(\mathbf{1})}_T(t_d)$ in region $II$ are also provided with vertical dashed yellow lines in Fig. \ref{Fig:Density_average}(b).
The lower one at $6.5 \, \rm{kHz}$ corresponds to the difference between the first excited Efimov state and the first atom-dimer~\cite{Bougas_interferometry_2023}.
A superposition of two such states was identified in Refs.~\cite{yudkin_coherent_2019,yudkin_reshaped_2024}, where the Ramsey-type interferometer has been employed for $^7$Li at repulsive interactions. 
The resulting oscillatory fringes then allowed the precise measurement of the binding energy of the Efimov state.
At the high frequency end of the spectrum, two pronounced peaks appear at $63.8 \, \rm{kHz}$ and $70.4 \, \rm{kHz}$.
These refer to $\abs{E^{(1)}_{AD}-E^{(1)}_A}/h$ and  $\abs{E^{(2)}_{T}-E^{(1)}_A}/h$ respectively [vertical dashed yellow lines in Fig. \ref{Fig:Density_average}(b)].
When analyzing the frequency spectrum in region $I$ though, a single pronounced peak appears, referring to $\abs{E^{(2)}_{T}-E^{(1)}_A}/h$.
It is therefore possible to measure the binding energy of Efimov trimers without requiring an atom-dimer as a reference state. This behavior grants the applicability of the interferometry protocol to attractive interactions as well, where atom-dimers are absent.

Turning to the frequency spectrum stemming from $\braket{ \mathbb{P}_T(t_d) }$ [blue solid line in Fig. \ref{Fig:Density_average}(b)], a low frequency peak appears also in the same position as for $ \mathbb{P}^{(\mathbf{1})}_T(t_d)$, i.e. $\abs{E^{(1)}_{AD}-E^{(2)}_T}/h$.
Essentially, the density profile of a semi-classical thermal gas does not induce any significant shift in the frequency related to the superposition of the Efimov trimer with the atom-dimer.
This outcome could contribute to understanding the 
impact of an interacting thermal gas
on the interference fringes~\cite{zhang_many_2023}.
The two high frequencies, however, are slightly shifted to lower values with respect to the spectra associated to the peak density [yellow vertical dashed lines in Fig. \ref{Fig:Density_average}(b)].
In fact, they are shifted to values close to $\abs{E^{(2)}_T}/h$ and $\abs{E^{(1)}_{AD}}/h$ [vertical black dash-dotted lines in Fig. \ref{Fig:Density_average}(b)].

Such frequencies correspond to the energy difference of the Efimov trimer and atom-dimer with the energy of the three-atom dissociation threshold in free space~\cite{naidon_efimov_2017}. Since the two former states are deeply bound, their energy has a weak dependence on the trap, assuming almost the same value in free space.
On the other hand, the energy of the first three-atom state highly depends on the trap frequency, and in the non interacting case it explicitly reads $3 \hbar \omega^{(\mathbf{j})}_r$.
Therefore, similarly to a thermal average~\cite{Bougas_interferometry_2023}, the density average in Eq. \eqref{Eq:Density_average_final} smears out all contributions stemming from three-atom states.
Eventually only oscillatory fringes survive, with frequencies associated to the energies of the trimer and first atom-dimer.
Since the two latter states are deeply bound, the above mentioned frequency shifts are small.
In order to observe such shifts, the $F(\omega)$ spectrum is shown in Fig. \ref{Fig:Density_average}(b) for both regions $I$ and $II$.
In region $I$, a single dominant peak arises due to the superposition of the first excited trimer and the first three-atom state pertaining to $\omega^{(1)}_r$, i.e. associated to the peak density $n_0$.
At such small dark times $t_d$, mostly high frequencies are contributing to the density average in Eq. \eqref{Eq:Density_average_final}, and thus the impact of lower trap frequencies is not appreciable in this region.

Subsequently, the dependence of the above identified shifts on the frequency  $\omega^{(\mathbf{c})}_r$ corresponding to the lowest density of the thermal gas is investigated.
This is carried out within the three-level model, which was shown to be an adequate qualitative description of the dynamical association processes in Ref.~\cite{Bougas_interferometry_2023}. 
Moreover, it can qualitatively capture the characteristics of the frequency spectrum presented in Fig. \ref{Fig:Density_average}(b).
The main premise of the model is to take into account only three states out of the entire spectrum [Fig. \ref{Fig:Spectrum}]; the first excited Efimov trimer, the first atom-dimer and the first three-atom state. 
The three possible superpositions of pairs of these states give rise to the three dominant frequencies in the spectra [vertical dashed yellow lines in Fig. \ref{Fig:Density_average}(b)].
Here, we extend the three-level model to three particles trapped with frequencies associated to densities lower than $n_0$.
Essentially, for all distinct $\omega^{(\mathbf{j})}_r$, only the three-atom states are significantly different in energy out of the three levels. 
The resulting $\mathbb{P}^{(\mathbf{j})}_T(t_d)$ are subsequently averaged according to Eq. \eqref{Eq:Density_average_final}.

The position of the high frequency peak $\omega/(2\pi)$ close to $70 \, \rm{kHz}$ stemming from the frequency spectrum of the extended three-level model is then investigated with respect to $\omega^{(\mathbf{c})}_r/\omega^{(\mathbf{1})}_r$ [blue solid line in Fig. \ref{Fig:Cutoff_dependence}]. 
The horizontal dash-dotted and dashed lines are associated to the energy gaps of the trimer with the first three-atom state ($\abs{E^{(2)}_T-E^{(1)}_A}/h$) and the trimer with the three-atom threshold ($\abs{E^{(2)}_T}/h$) respectively. 
These values act as a reference for identifying changes in the peak position. 
When $\omega^{(\mathbf{c})}_r/\omega^{(\mathbf{1})}_r \simeq 1$, a frequency similar to $\abs{E^{(2)}_T-E^{(1)}_A}/h$ is identified [dash-dotted line in Fig. \ref{Fig:Cutoff_dependence}].
As $\omega^{(\mathbf{c})}_r$ further diminishes, $\omega/(2\pi)$ decreases as well in a step-wise manner.
This behavior is due to the finite resolution of the Fourier transform, which cannot capture small gradual frequency changes for the final dark times considered here. 
For $\omega^{(\mathbf{c})}_r/\omega^{(\mathbf{1})}_r  \lesssim 0.7$, the frequency reaches a plateau below $\abs{E^{(2)}_T-E^{(1)}_A}/h$, but above the frequency associated to just the energy of the Efimov trimer [dashed line in Fig. \ref{Fig:Cutoff_dependence}]. 
This suggests that as the lowest density dependent frequency $\omega^{(\mathbf{c})}_r$ further decreases past this threshold, the high-frequency peak shifts towards the value $\abs{E^{(2)}_T}/h$.

\section{Conclusions}  \label{Sec:Conclusions}

We have addressed the impact of the density profile of a thermal gas on the dynamical association of Efimov trimers by means of a Ramsey-type protocol~\cite{yudkin_coherent_2019,yudkin_reshaped_2024}. 
To this end, the time-dependent three-body problem is solved for an ensemble of density dependent trapping frequencies. 
The latter correspond to a range of density values, spanning the density profile of a thermal gas.
In consistence with the results observed exclusively for the peak density~\cite{Bougas_interferometry_2023}, three dominant frequencies appear in the frequency spectrum of the normalized probability to occupy Efimov trimers.
The lowest one corresponds to the superposition of the first excited Efimov trimer with the first atom-dimer, being the same as the one stemming only from the peak density. 
Essentially, the density profile of a thermal gas does not affect the frequency associated to this superposition, which was instrumental in measuring the binding energy of Efimov trimers~\cite{yudkin_coherent_2019,yudkin_reshaped_2024}.
On the other hand, the two high frequencies are slightly shifted from the values pertaining only to the peak density, i.e. the energy difference between the trimer and the first three-atom state, and between the latter and the first atom-dimer.
It is shown that the high frequencies are slightly shifted to lower values, towards the energies of the trimer and the first atom-dimer in free space. 
Such a behavior is attributed to the smearing effect that the density average has on the highly-oscillatory fringes associated with three-atom states.

A further question that arises from our study is whether the density profile of a thermal gas will affect the decay times of the oscillatory fringes due to the lifetime of Efimov states. 
Concretely, whether the decay times will be twice as long as the lifetime of trimers, a result established only in the case where the peak density is considered~\cite{Bougas_interferometry_2023}. 
Moreover, an important future aspect of our work is to address the impact of the medium 
on Efimov states. 
Such a study requires not only the consideration of the density profile, but also the role of interactions and correlations of the host gas.

\bmhead{Acknowledgements}

    We acknowledge useful discussions with J. P. D'Incao.

    \section*{Declarations}

\begin{itemize}
\item Conflict of interest/Competing interests The authors declare no conflict of interest.
\end{itemize}

\bibliography{few_body_combined_2}

\end{document}